\documentclass[12pt]{article}

\usepackage{epsfig}
\usepackage{amssymb}

\def\lsim{\mathrel{\rlap{\lower4pt\hbox{\hskip1pt$\sim$}}
    \raise1pt\hbox{$<$}}}         
\def\gsim{\mathrel{\rlap{\lower4pt\hbox{\hskip1pt$\sim$}}
    \raise1pt\hbox{$>$}}}         

\def\overleftrightarrow#1{\vbox{\ialign{##\crcr
    $\leftrightarrow$\crcr
    \noalign{\kern 1pt\nointerlineskip}
    $\hfil\displaystyle{#1}\hfil$\crcr}}}

\newcommand {\be}{\begin{equation}}
\newcommand {\ee}{\end {equation}}

\begin{document}

\hspace{12cm}{\bf TUM/T39-00-22}\\

\hspace{12cm}{\bf NT@UW-00-033}\\

\begin{center}
{\bf Coordinate Space 
Distributions of Antiquark Flavor Asymmetries\\ in the Proton
\footnote{work supported in part by BMBF, by the Alexander von
Humboldt Foundation and by the US Dept. of Energy}}
\end{center}
\begin{center}
{Ernest M. Henley$^{a,b}$, Thorsten Renk$^{c}$, and Wolfram Weise$^{c}$} 

{\small \em $^{a}$ Department of Physics, University of Washington, Seattle,
WA 98195, USA} \\
{\small \em $^{b}$ Institute for Nuclear Theory, University of Washington, 
Seattle, WA 98195, USA }\\
{\small \em $^{c}$ Physik Department, Technische Universit\"{a}t M\"{u}nchen, 
D85747 Garching, GERMANY}
\end{center}
\vspace{0.25 in}
\begin{abstract}

We examine the space-time  properties of the distributions of 
$(\bar {d} - \bar{u})$ and $
\bar{d}/\bar{u}$ in the proton. The difference of the antiquark distributions 
shows the expected peak at the approximate pion Compton wavelength and
is supportive of the thesis
that the meson cloud of the nucleon is at the origin of the asymmetry of
$(\bar {d} - \bar{u})$, with the pion cloud playing a dominant role. 
\end{abstract}

\vspace {0.25 in}

The partonic distribution functions of the proton,
particularly the flavor dependence of the antiquark distributions, remain of 
considerable interest. Experimentally, the 
NMC measurement of the integral of
$(\bar{d} - \bar{u})$ \cite{NMC}, and the more recent 
measurements of the ratio $\bar{d}/\bar{u}$ by means of the Drell-Yan
process \cite {NA51,E866} provide evidence for the excess of $\bar{d}$ over
$\bar{u}$ in the proton. Similar results 
for $(\bar{d} - \bar{u})$ were obtained by HERMES
\cite{hermes}. One of the simplest (physically "anschaulich") explanations 
for the excess is based on the meson cloud model 
and the Sullivan process. This model can explain the 
momentum fraction (Bjorken x-dependence) of the
$(\bar{d} - \bar{u})$ distribution (for  reviews, see
\cite{S&T}).  Pions play the leading
role in this context, while the detailed description
of the ratio $\bar{d}/\bar{u}$ may also require correlated
$q\bar{q}$ pairs of heavier mass\cite{E866,AH,AHM}.

In this Letter we do not wish to dwell on the differences between 
theoretical models and experiment, but examine the spacetime 
properties of the flavor distribution asymmetries in order to see whether 
they offer any clues as to their origin. 
In order to obtain the distribution functions in 
coordinate space, we follow Piller et al. \cite {Pil} and
 V\"{a}nttinen et al. \cite {Va}.
We work with light-cone variables and
introduce the dimensionless coordinate spacetime variable $z =
y \cdot P$, where $P$ is the momentum of the nucleon. 
 The 
light cone distance is $y^+ \equiv t + y_3 =  2z/M$, with $M$
the nucleon mass. The dimensionless space variable
$z$ is conjugate to Bjorken $x$ and $z \simeq 5$ corresponds to
$y^+ \simeq 2$ fm or a longitudinal distance of approximately 1 fm. 

In the first instance we use empirical 
distribution functions and differences thereof. 
In accordance with the charge conjugation properties
of momentum space quark distributions we find the coordinate space
distribution of the sea quarks as

\begin{equation}
Q_{sea}(z,Q^2) = \int_0^1 dx \left[q_{sea}(x,Q^2) +\bar{q}(x,Q^2) \right]\sin(zx)   
\label {1}
\end{equation}

\noindent
Decomposing this result into the contributions from quarks and
antiquarks of different flavours we arrive at an expression for the
asymmerty in coordiante space: 

\begin{equation}
(\bar{D} - \bar{U})(z,Q^2)= \int_0^1 dx \left[ \bar{d}(x,Q^2) - \bar{u}(x,Q^2)\right]\sin(zx)
\label {1a}
\end{equation}

The distribution functions $\bar{d}(x), \bar{u}(x)$ have been obtained from 
various deep inelastic electron scattering scattering and Drell-Yan 
experiments by CTEQ5 and other groups \cite{CTEQ5,MRST,GRV}. In Fig.1 we 
show $(\bar{d} - \bar{u})$ and $(\bar{D} - \bar{U})$ 
obtained from Eq.(\ref{1a}) and the CTEQ5, the MRST and the GRV analysis.
We observe that all three parameterizations agree fairly well between 
each other except for
small values of $x$ or large $z$. 
Therefore we will mainly use the results from CTEQ5 in the following
for comparison to the pion cloud model.


\begin{figure}[!htb]
\begin{center}
\epsfig{file=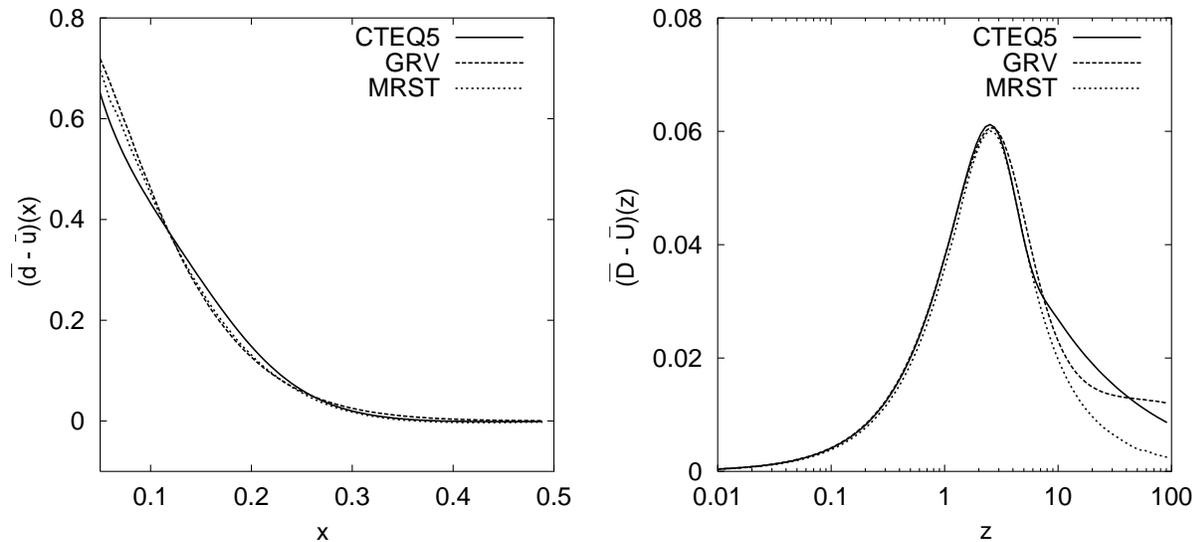, width=\textwidth}
\end{center}
\caption{Momentum space and coordinate space distribution of the
$\bar{d} -\bar{u}$ asymmetry in the proton at $Q^2 = 4$ GeV$^2$, based on the
parameterizations \cite{CTEQ5,MRST,GRV}.}
\end{figure}


Investigating the coordinate space distribution, we see that
for $Q^2 = 4 \,$GeV$^2$    
the peak of $(\bar{D} - \bar{U})$
occurs at $z \simeq 3$ or at $y^+ \simeq 1.2 $ fm. The half-width of
the peak extends from $ z \simeq 1 - 10$ or $ y^+ \simeq 0.4 -4 $ fm. 
This is a region where the pion and perhaps more massive $q\bar{q}$ states \cite{AHM}
contribute. We recall that at 3-4 fm, the valence quark distribution has fallen to 
less than  25\%  of its peak value.

\begin{figure}[htb]
\begin{center}
\epsfig{file=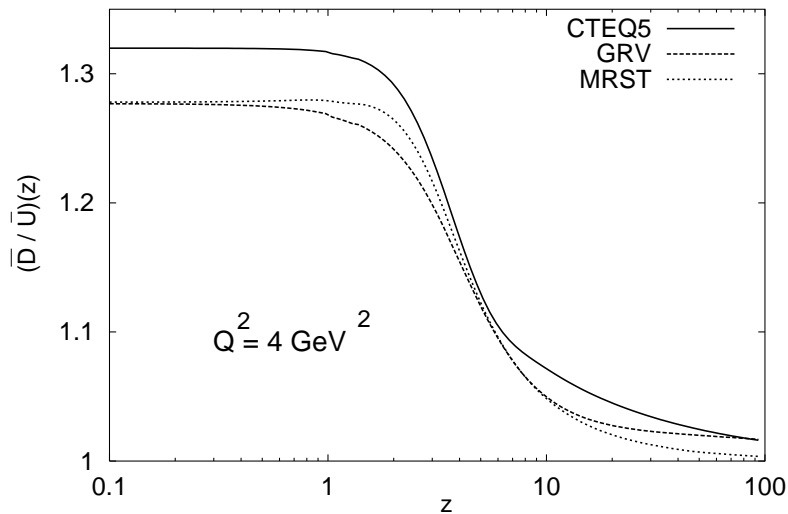,width=10.5cm}
\end{center}
\caption{\label{DoverU}The Ratio $ R(z)=\bar{D}(z) / \bar{U}(z)$ in coordinate space
for different parametrizations.}
\end{figure}

In Fig.~\ref{DoverU} we show the ratio $R(z)= \bar{D}(z) / \bar{U}(z)$ as a function of
$z$. Here the small z region is expected to be constant because 
$\bar{q}(x) \simeq 0$ for $x \geq 0.35$, so that for $z \leq 1$, Eq. 
(\ref{1})
becomes

\begin{equation}
\bar{Q}(z) \simeq  z \int_0^1 \bar{q}(x) \,x\; dx = constant \cdot z \; .
\end{equation}

\noindent
It follows that the ratio $R(z) $ is a constant for $z \lsim 1$. Most of the "action"
takes place where the derivative of $R(z)$ differs appreciably from zero, which 
again occurs for $ z \simeq 2-10$, corresponding to distance scales
comparable to the size of the nucleon. We interpret these features as being 
supportive of a meson cloud as the primary cause for the excess of 
$\bar{d}(x)$ over $\bar{u}(x)$ in the proton.


In order to make a detailed comparison with the meson cloud model and 
Sullivan process, we have used it in its simplest form with only pions and 
no $\Delta$ and have concentrated on $(\bar{d} - \bar{u})$ because the 
detailed description of the 
ratio $\bar{d}/ \bar{u}$ also requires the perturbative contribution 
from gluon splitting which is expected to be symmetric in $\bar{d}$ and
$\bar{u}$. 

We provide the usual formulae \cite{S&T} for the effects of the 
meson cloud (and Sullivan process). 
The wave function of the proton is
written in terms of Fock states with and without mesons:

\begin{eqnarray}
\mid p\rangle = \sqrt{Z}\mid p\rangle_{\rm bare} + \sum_{MB}\int dy\;
d^2 k_\perp\; \phi_{BM}(y,k_\perp^2)
\mid  B(y,\vec{k}_\perp) M(1-y, - \vec{k}_\perp)\rangle \;.
\end{eqnarray}

\noindent
Here $\sqrt{Z}$ is a wavefunction renormalization constant,  $\phi_{BM}(y,k_\perp^2) $
is the probability amplitude for finding a physical nucleon in a state
consisting of a baryon, $B$, with longitudinal momentum fraction y, and a meson,
$M$, of 
momentum fraction $(1-y)$ and squared transverse relative momentum $k_\perp^2$.

The quark distribution function $q(x)$ of a proton is given by

\begin{equation}
q(x) = q^{\rm bare}(x) + \delta q(x) \; ,
\end{equation}

\noindent
with 

\begin {eqnarray}
\delta q(x) = \sum_{MB}
\left(\int_x^1 f_{MB}(y) q_M \left(\frac{x}{y}\right)\frac{dy}{y} \; 
+\int_x^1 f_{BM}(y) q_B \left(\frac{x}{y} \right) \frac{dy}{y}\right),
\end{eqnarray}

\noindent
where $q_M$ and $q_B$ are the quark distributions in the meson and baryon,

\begin{equation}
f_{MB}(y) = f_{BM}(1-y) \; ,\label{sym}
\end{equation}

\begin{equation}
f_{BM}(y) = \int_0^\infty\mid\phi_{BM}(y,k_\perp^2)\mid^2~d^2k_\perp\;.
\end{equation}

\noindent
The meson--baryon vertex function $\phi_{BM}$ includes a cutoff factor

\begin{equation}
G_M (t,u) = \exp{\frac {t- m_M^2}{2 \Lambda_M^2}}
\exp{{ u - m_B^2\over 2\Lambda_M^2}} \; ,
\end{equation}

\noindent
where $\Lambda_M$ is a cut-off parameter for pions and $t$ and $u$
are the usual Mandelstam kinematical
variables, expressed in terms of $\vec{ k}_\perp$ and $y$.
Such a form is required to respect the identity
(\ref{sym})\cite{S&T, Szczurek:1996ur}.

The cut-off required in the model is taken from ref. \cite {S&T}, but 
it is also varied to study its effect. We found that the cutoff basically
regulates the overall normalization of the result and used the value
which leads to the best fit ($\Lambda_M$ = 0.85 GeV).

The expressions for the
splitting functions  $f_{MB}(y)$ are those  given by \cite {S&T}
as derived in 
Ref.~\cite{holtmann}.
We include only the pion; then the renormalization constant of the
"bare" quark is $Z = 1-3 n_\pi$ where $n_\pi$ is the
probability to find a neutral pion in the cloud.

We need the valence quark distributions in the pion \cite{sutton},
$q_M(x)$, which is given at $Q^2 = 4$ GeV $^2$ as

\begin{eqnarray}
x q_M(x)=0.99x^{0.61}(1-x)^{1.02},
\end{eqnarray}

\noindent
and its $Q^2$ evolution.
At this point, a problem arises  caused by the fact that the parton 
distributions of the pion were extracted from $\pi N$ scattering assuming
that the nucleon ones are known.
Since in our approach the nucleon also has an admixture of the pion cloud, 
the extraction of the
pionic parton distributions is not fully consistent within this framework
and therefore we do not expect a perfect agreement with data in the end.
We do not need the 
bare nucleon sea since it averages out in the difference 
$(\bar{d} - \bar{u})$, which is our main interest.

\begin{figure}[htb]
\begin{center}
\epsfig{file=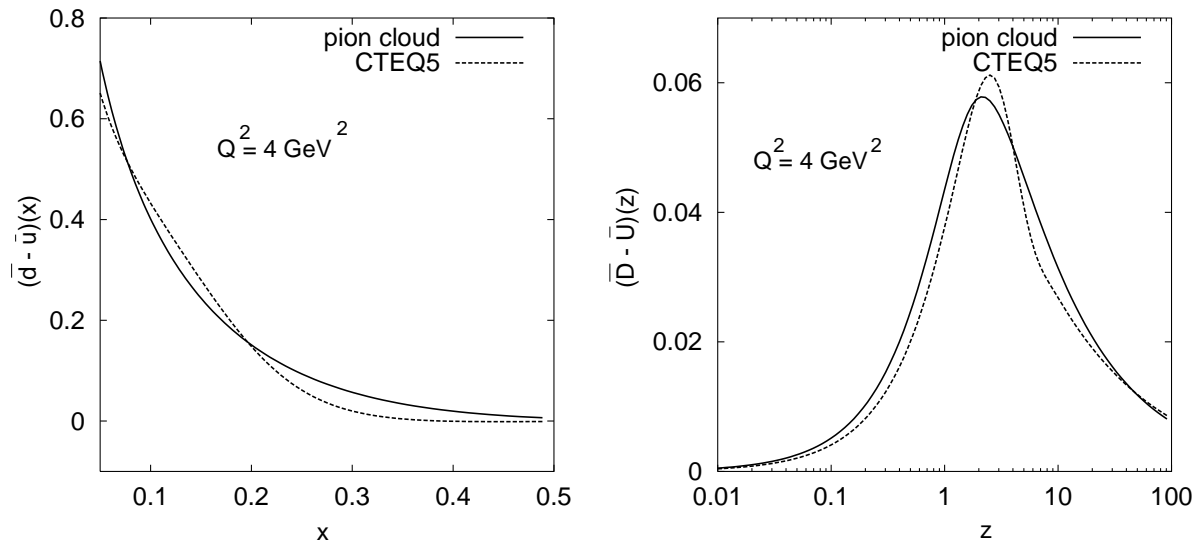,width=\textwidth}
\end{center}
\caption{\label{PCvsCTEQ}The asymmetry $\bar{d}-\bar{u}$ in momentum space and
$\bar{D}-\bar{U}$ in coordinate space in the pion cloud model calculations
and in the CTEQ5 parametrization.}
\end{figure}

The results of our calculations for $\bar{d}(x)-\bar{u}(x)$,
using a cutoff $\Lambda_M = 0.85$ GeV,
are compared to CTEQ5
in Fig.~\ref{PCvsCTEQ}.


\noindent
Also shown in this figure is the space coordinate transform $\bar{D} - \bar{U}$.
The peak occurs for a somewhat lower z than obtained by CTEQ5,
however the overall agreement is obvious.

\begin{figure}[!htb]
\begin{center}
\epsfig{file=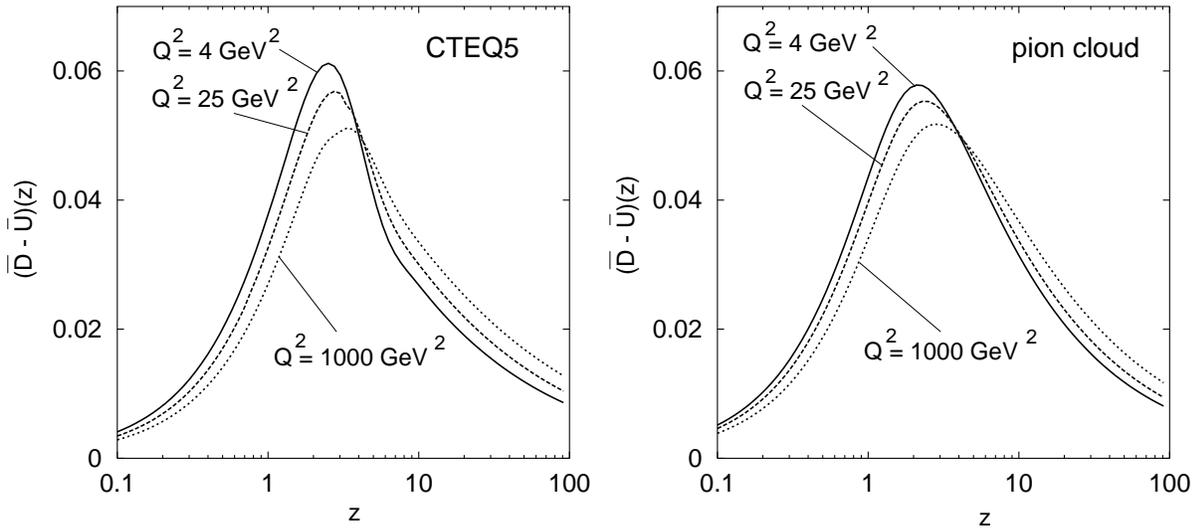,width=\textwidth}
\end{center}
\caption{\label{Evolution}
$Q^2$-evolution of $\bar{D} -\bar{U}$ in the pion cloud model
and in the CTEQ5 parameterization.}
\end{figure}

Finally, Fig.~\ref{Evolution} shows the $Q^2$ evolution of the results obtained
within the pion cloud model and in the CTEQ5 asymmetry. 
The $Q^2$ dependence of the pion cloud asymmetry evidently has 
the same qualitative feature as
in the $Q^2$ evolution of the CTEQ5 results, namely a shift of
the peak to higher $z$ and a decrease of the maximum peak value
with increasing $Q^2$.


We summarize and conclude with the following observations. The spacetime coordinate
representation of sea quark distributions offers detailed insights, additional to 
their momentum space form, into the flavour asymmetry $\bar{d}-\bar{u}$ of
antiquarks. In the $Q^2$ range between 5 and 25 GeV$^2$,
this asymmetry is maximal in coordinate space at length scales (1.2 -- 1.6) fm
characteristic of the pion cloud of the nucleon. These features are
evident in the empirical asymmetry distributions, and they
are well undersood in the pion cloud model.
Additional contributions of heavier masses may assist
in reproducing the detailed behaviour of the $\bar{d}/\bar{u}$ ratio.

\vspace{3ex}

We thank Gunther Piller and Lech Mankiewicz for helpful
comments and discussions.

\end{document}